\documentclass{aa}  

\usepackage{graphicx}
\usepackage{txfonts}
\usepackage{natbib} 
\bibpunct{(}{)}{;}{a}{}{,} 

\usepackage{verbatim} 
\usepackage{tikz}
\usetikzlibrary{positioning}
\usepackage{hyperref}   
\hypersetup{colorlinks=true,linkcolor=blue,citecolor=blue,filecolor=blue,urlcolor=blue}

\begin{document}

   \title{Age and convective core overshooting calibrations in CPD-54 810 binary system} 

   \subtitle{Statistical investigation on the solution robustness}
  \author{G. Valle \inst{1, 2}, M. Dell'Omodarme \inst{1}, P.G. Prada Moroni
        \inst{1,2}, S. Degl'Innocenti \inst{1,2} 
}
\titlerunning{CPD-54 810 binary system}
\authorrunning{Valle, G. et al.}

\institute{
        Dipartimento di Fisica "Enrico Fermi'',
        Universit\`a di Pisa, Largo Pontecorvo 3, I-56127, Pisa, Italy
        \and
        INFN,
        Sezione di Pisa, Largo Pontecorvo 3, I-56127, Pisa, Italy
}

   \offprints{G. Valle, valle@df.unipi.it}

   \date{Received 11/03/2023; accepted }

  \abstract
{}
{Relying on recent very precise observations for the CPD-54 810 binary system, we investigate the robustness of the estimated age and convective core overshooting for a system with both stars on the main sequence (MS). Our main aim is to assess the variability in the results, accounting for different statistical and systematic sources of uncertainty.}
{We adopt the SCEPtER pipeline, a well-established maximum likelihood technique,  based on fine grids of stellar models computed for a different initial chemical composition and convective core overshooting efficiency.}
{We performed different fits of the system, under different assumptions. The base fit suggests a common age of $3.02 \pm 0.15$ Gyr, in agreement with recent literature.     
This estimated convective core overshooting parameter is $\beta = 0.09 \pm 0.01$, with a corresponding convective core
mass $M_c = 0.059^{+0.017}_{-0.021}$~$M_{\sun}$.
The robustness of these estimates were tested assuming a narrow constraint on the helium-to-metal enrichment ratio, which is in agreement with the recently published results on the Hyades cluster. Under this constraint the chemical solution of the system changes, but the age and the overshooting parameter are almost unchanged ($3.08^{+0.17}_{-0.14}$ Gyr and $0.09 \pm 0.01$). In a further test, we halved the uncertainty as to the effective temperature of both stars and again the estimated parameter shows only small variations ($3.02 \pm 0.12$ Gyr and $0.09 \pm 0.01$).}
{This low variability suggests that the age of the system with both stars in the MS can be reliably estimated at a 5\% level, but it also indicates that the power of the investigation is probably low, because it is possible to find a satisfactory fit in several different configurations by only varying the initial chemical composition within its uncertainty. Despite the great increase in the observational constraints' precision, the results support the conclusions of previous theoretical works on the stellar parameter calibration with double MS star binary systems. } 

   \maketitle

\section{Introduction}\label{sec:intro}

While stellar model predictions have become very accurate in the last decades, they are still nonetheless affected by some weakness, one of the major ones being the lack of a rigorous treatment of
convective transport \citep[see][for a comprehensive introduction]{Viallet2015}.
This limitation hampers a firm prediction of the dimension of the convective core. Stellar modellers usually compute its extension beside the classical Schwarzschild convective region allowing for an overshooting zone, whose extension is a function of a free parameter to be empirically calibrated. 
However, it has been shown in the literature \citep{Claret2017, binary, overshooting, TZFor} that only very precise observations of binary systems could, in principle, be used for model calibration.  

The outstanding improvement in the precision of radii and masses for stars in a binary system -- thanks to satellite missions such as Kepler and TESS \citep{Borucki2010, Ricker2015} --  coupled with the advancements made to the ways in which methods estimate  the stellar effective temperatures \citep{Miller2020}, nowadays allow for a precision in the measured parameters below 1\% to be achieved for well-studied targets. This relevant improvement might modify the conclusions of studies even from the recent pasts that adopted much larger uncertainties in their computations \citep[e.g.][]{overshooting}.

An ideal target to investigate the robustness and variability of fundamental parameter estimates was recently identified in the CPD-54 810 binary system, which was extensively analysed by \citet{Miller2022}, who adopted a powerful technique to estimate the stellar effective temperatures, relying on estimates of masses and radii at the 0.1\% level.
The system is composed of two stars on the main-sequence (MS) evolutionary stage and is therefore a perfect target to investigate the possible changes in the conclusions of \citet{overshooting} -- who advised against adopting targets in the MS for calibration purposes -- thanks to the improvement  of an order of magnitude in the observational uncertainties.

A preliminary attempt to constrain the age of the system, profiting from the achieved precision, was performed by \citet{Miller2022}. However, the focus of that paper was not on the stellar evolution calibration. Here we specifically address this topic by investigating different scenarios.  
In fact, the main interest in performing a fit of such a system, besides the obvious interest in the age estimation, is to establish the robustness of the fitted stellar parameters. 

Therefore, in this paper, we investigate several systematic effects that can bias the results, owing to different but legitimate decisions of the researchers in the fitting process. It should, however, be noted that we perform our analysis fixing the input physics of the adopted theoretical stellar models. A non-negligible variability is expected when results from different pipelines are compared on common targets \citep[see e.g.][]{Reese2016, Stancliffe2016,  SilvaAguirre2017, TZFor}. Therefore, our results can be considered as representative of the random uncertainty achievable with a single pipeline approach. In the concluding section, we further discuss what is expected from a multi-pipeline analysis.

The structure of the paper is as follows. In Sect.~\ref{sec:method}, we 
discuss the method and the grids used in the estimation process. 
The  fit of the system is
 presented in Sect.~\ref{sec:results}, with an analysis of the impact of different assumptions in the adopted grids and observational uncertainties.
Some concluding remarks and a comparison with the literature can be found in Sect.~\ref{sec:conclusions}.

\section{Methods and observational constraints}\label{sec:method}

\subsection{Fitting technique}

The analysis was conducted adopting the SCEPtER pipeline\footnote{Publicly available on CRAN: \url{http://CRAN.R-project.org/package=SCEPtER}, \url{http://CRAN.R-project.org/package=SCEPtERbinary}}. This technique is well tested and adopted in the literature for single stars and binary systems 
\citep[e.g.][]{scepter1,eta,bulge, binary, TZFor}. 
The procedure provides estimates of the parameters of interest (age, initial helium abundance, initial metallicity, core overshooting parameter,  and extension of the convective core) adopting a maximum likelihood over a grid approach.

The method is explained in detail in \citet{binary}; here, we provide only a brief summary for the reader's convenience. Basically, for every $j$-th point in the estimation grid of pre-computed stellar models, a likelihood estimate was obtained for both stars
\begin{equation}
        {{\cal L}^{1,2}}_j = \left( \prod_{i=1}^n \frac{1}{\sqrt{2 \pi}
                \sigma_i} \right) 
        \times \exp \left( -\frac{\chi^2}{2} \right)
        \label{eq:lik}
        ,\end{equation}
\begin{equation}
        \chi^2 = \sum_{i=1}^n \left( \frac{o_i -
                g_i^j}{\sigma_i} \right)^2
        \label{eq:chi2},
\end{equation}
where $o_i$ are the $n$ observational constraints, $g_i^j$ the $j$-th grid point corresponding values, and $\sigma_i$ the observational uncertainties.

Then, the joint likelihood of the system was computed as the product of the single star likelihood functions.  
It is possible to obtain the estimates for both the two individual components and for 
the  whole system. In the former case, the fits for the two stars were obtained independently,
while in the latter case the two objects must have a common age (with a tolerance of 1 Myr), as well as an identical initial helium abundance and initial metallicity.

The error on the estimated parameters was obtained by means of Monte Carlo simulations. 
We generated $N = 10\,000$ artificial binary systems, sampling from a multi-variate Gaussian distribution centred on the observational data, taking into account the correlation structure among the two star observational data. 
As in \citet{TZFor}, we assumed a correlation of 0.95 between
the primary and secondary effective temperatures, and 0.95
between the metallicities of the two stars. Regarding mass and radius correlations, the high precision of the estimates makes these parameters of no importance, but we set it at 0.8 for the masses and -0.9 for the radii, which are typical values for this class of stars \citep{binary, TZFor}. We explicitly verified that different adoptions of masses and the correlation of radii lead to negligible modifications in the results. 

\subsection{Observational data}

\begin{table}
        \centering
        \caption{Observational constraints for the CPD binary system from \citet{Miller2022}, with an increased uncertainty as to the effective temperatures.}
        \label{tab:input}
        \begin{tabular}{lcc}
                \hline\hline
                & primary & secondary \\
                \hline 
                $M$ ($M_{\sun}$) & $1.3094 \pm 0.0051$ & $1.0896 \pm 0.0034$ \\
                $R$ ($R_{\sun}$) & $1.9288 \pm 0.0030$ & $1.1815 \pm 0.0037$ \\
                $T_{\rm eff}$ (K) &  $6462 \pm 100$ &  $6331 \pm 100$ \\
                ${\rm [Fe/H]}$ & $0.0 \pm 0.2$ & $0.0 \pm 0.2$\\
                \hline
        \end{tabular}
\end{table}

As observational constraints,  we used the masses, radii, metallicities [Fe/H], and effective temperatures of both stars. The adopted values and their uncertainties, reported in Table~\ref{tab:input}, are taken from \citet{Miller2022}.

The uncertainty as to the effective temperature reported in \citet{Miller2022} is 43 K; however, as it is discussed extensively in that paper, there are possible systematic effects that might modify the calibration scale. Therefore, as a reference scenario, we assumed a prudential estimate of 100 K as an error in $T_{\rm eff}$ for both stars. We discuss in Sect.~\ref{sec:teff} the modification of the parameter estimates assuming a more precise effective temperature constraint.

\subsection{Stellar models' grid}

The grids of models were computed for the  mass in the ranges [1.08, 1.10] $M_{\sun}$ and [1.30, 1.32] $M_{\sun}$ with a step of 0.002 $M_{\sun}$, from the pre-MS up to the start of the red giant branch (RGB). 
The initial metallicity [Fe/H] was varied from $-0.4$ dex to 0.3 dex, with
a step of 0.05 dex. 
The solar heavy-element mixture by \citet{AGSS09} was adopted\footnote{A reduced test was conducted adopting the \citet{GS98} heavy-element mixture, with negligible differences in the results.}. 
Several initial helium abundances were considered at a fixed metallicity by adopting the commonly used
linear relation $Y = Y_p+\frac{\Delta Y}{\Delta Z} Z$
with the primordial abundance $Y_p = 0.2485$ from WMAP
\citep{peimbert07a,peimbert07b}. 
The helium-to-metal enrichment ratio $\Delta Y/\Delta Z$ was varied
from 1.45 to 2.55 with a step of 0.15, centred around 2.0 \citep{Tognelli2021}. 

Models were computed with the FRANEC code, in the same
configuration as was adopted to compute the Pisa Stellar
Evolution Data Base\footnote{\url{http://astro.df.unipi.it/stellar-models/}} 
for low-mass stars \citep{database2012}. 
The models were computed
by assuming the solar-scaled mixing-length parameter $\alpha_{\rm
        ml} = 1.74$.
The extension of the extra-mixing region beyond the Schwarzschild border
was considered only for the primary star and was parametrised  in terms of the pressure scale height $H_{\rm 
        p}$: $l_{\rm ov} = \beta H_{\rm p}$, with 
$\beta$ in the range
[0.00; 0.28] with a step of 0.01. The code adopts an instantaneous mixing in the overshooting treatment. Atmospheric models by \citet{brott05}, which were computed using the PHOENIX code
\citep{hauschildt99,hauschildt03} and are  
available in the range $3000 \; {\rm K} \le T_{\rm eff} \le 10000 \;
{\rm K}$, $0.0 \le \log g \; {\rm (cm \; s^{-2})} \le 5.0$, and $-4.0 \le {\rm
        [M/H]} 
\le 0.5$, were adopted.
In the range $10000 \; {\rm K} \le T_{\rm eff} \le 50000 \; {\rm K}$, $0.0
\le \log g \; {\rm (cm \; s^{-2})} \le 5.0$, and $-2.5 \le {\rm [M/H]} 
\le 0.5$, where models from \citet{brott05} are unavailable,    
models by \citet{castelli03} were used.
Further details on the stellar models are fully described in \citet{eta,binary} and the references therein.

\section{Results}\label{sec:results}

\subsection{Single stars and overall system fit}

\begin{table}
        \centering
        \caption{Results of the CPD-54 810 simultaneous binary system fitting.}
        \label{tab:fit-bin}
        \begin{tabular}{lccc}
                \hline\hline
                &  $q_{16}$ & $q_{50}$ & $q_{84}$ \\
                \hline 
                $Y$ & 0.268 & 0.280 & 0.289 \\
                $Z$ & 0.0103 & 0.0128 & 0.0158 \\
                $\beta$ & 0.080 & 0.090 & 0.105 \\
                $M_c$ ($M_\sun$) & 0.038 & 0.059 & 0.076 \\
                age (Gyr) & 2.87 & 3.02 & 3.17 \\
                \hline
                \multicolumn{4}{c}{Fit parameters}\\
                \hline
                $T_{\rm eff,1}$ (K) &  & 6449 &  \\ 
                $T_{\rm eff,2}$ (K) &  & 6251 &  \\             
                $M_1$ ($M_{\sun}$) &  & 1.310 &   \\ 
                $M_2$ ($M_{\sun}$) &  & 1.090 &  \\ 
                $R_1$ ($R_{\sun}$) &  &  1.928 &   \\ 
                $R_2$ ($R_{\sun}$) &  & 1.183 &  \\ 
                ${\rm [Fe/H]}_1$ &  & $-0.13$ & \\
                ${\rm [Fe/H]}_2$ &  & $-0.12$ & \\      
                \hline                
                $\chi^2$ &  & 1.6 &  \\ 
                \hline
        \end{tabular}
\end{table}

\begin{figure*}
        \centering
        \includegraphics[width=8.cm,angle=-90]{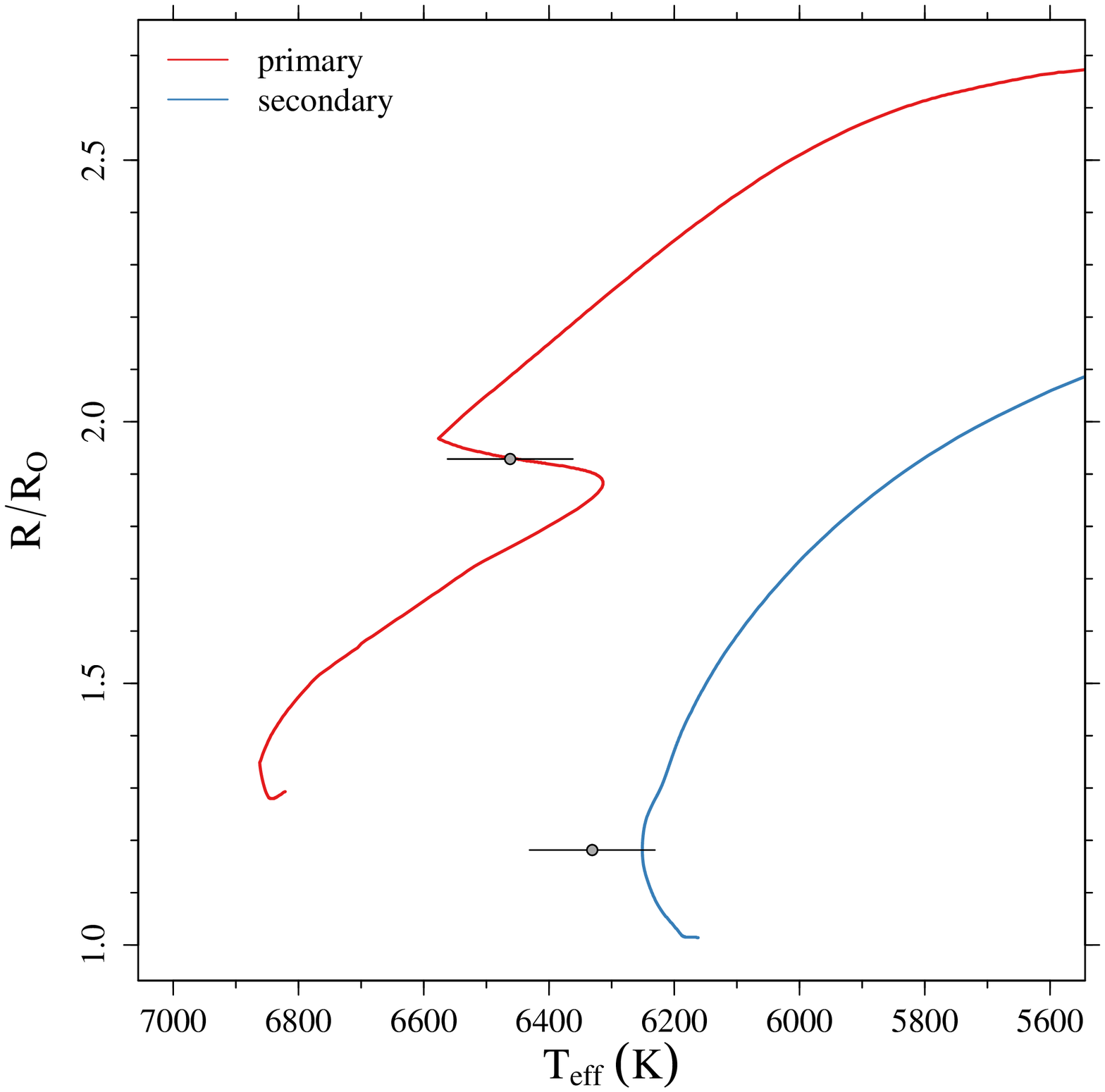}
        \includegraphics[width=8.cm,angle=-90]{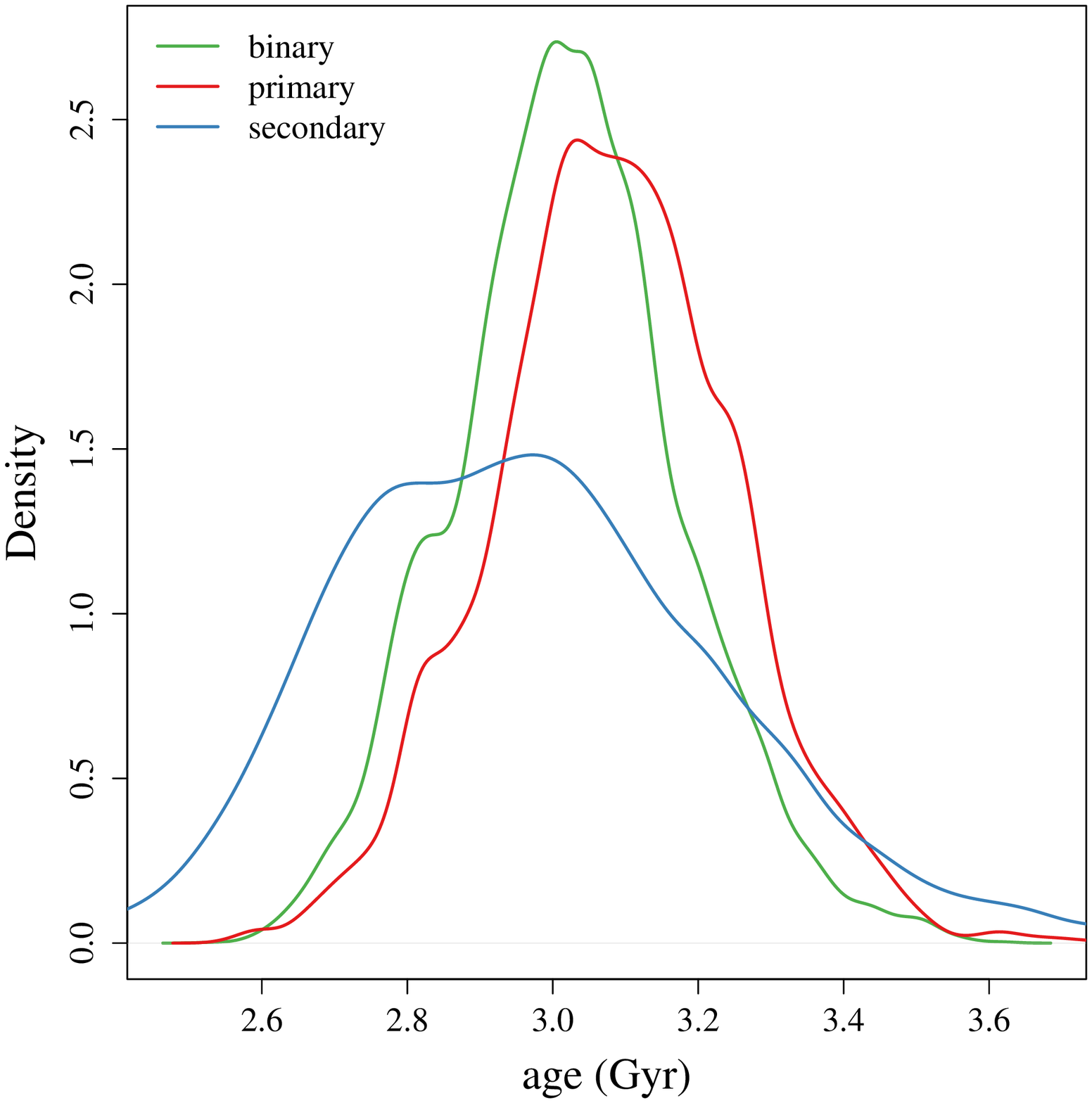}  
        \caption{Evolutionary tracks and core overshooting parameter from the fit of the binary system. {\it Left}: Comparison between the observational values of the effective temperature and radius of the two stars (grey circle) and the evolutionary tracks for the best solutions found in the analysis.  The error bars correspond to 1 $\sigma$ errors. {\it Right}: Density of probability for the estimated age of the system, and for the two individual stars.}
        \label{fig:HR}
\end{figure*}

The results of the estimation procedure applied to the whole system and to the individual stars are reported in Table~\ref{tab:fit-bin} and \ref{tab:fit-sing}, respectively.
The simultaneous fit of both stars of the system suggests a common age of $3.02 \pm 0.15$ Gyr, with an initial [Fe/H] = 0.0 and an initial helium abundance $Y = 0.28$. Therefore, a  helium-to-metal enrichment ratio  $\Delta Y /\Delta Z = 2.5^{+0.1}_{-0.65}$ is preferred. The position of the data with respect to the best fit evolutionary tracks is displayed in  Fig.~\ref{fig:HR} (left panel) in the radius versus effective temperature plane. The figure shows good agreement between theoretical models and observational data (the corresponding best fit models are not shown to improve the readability of the figure, and they correspond to the points where the error bars cross the evolutionary tracks). Qualitatively, the fit agrees with the one proposed by \citet{Miller2022}, with a primary star just above the hook in the overall contraction phase.  
To perform a formal assessment of the goodness of fit, we evaluated $\chi^2$ considering the differences between the fit values and the observational constraints, weighted by the observational uncertainties. We note that $\chi^2 = 1.6$ was obtained for eight variables and four parameters, suggesting a good fit.

\begin{figure*}
        \centering
        \includegraphics[width=8.cm,angle=-90]{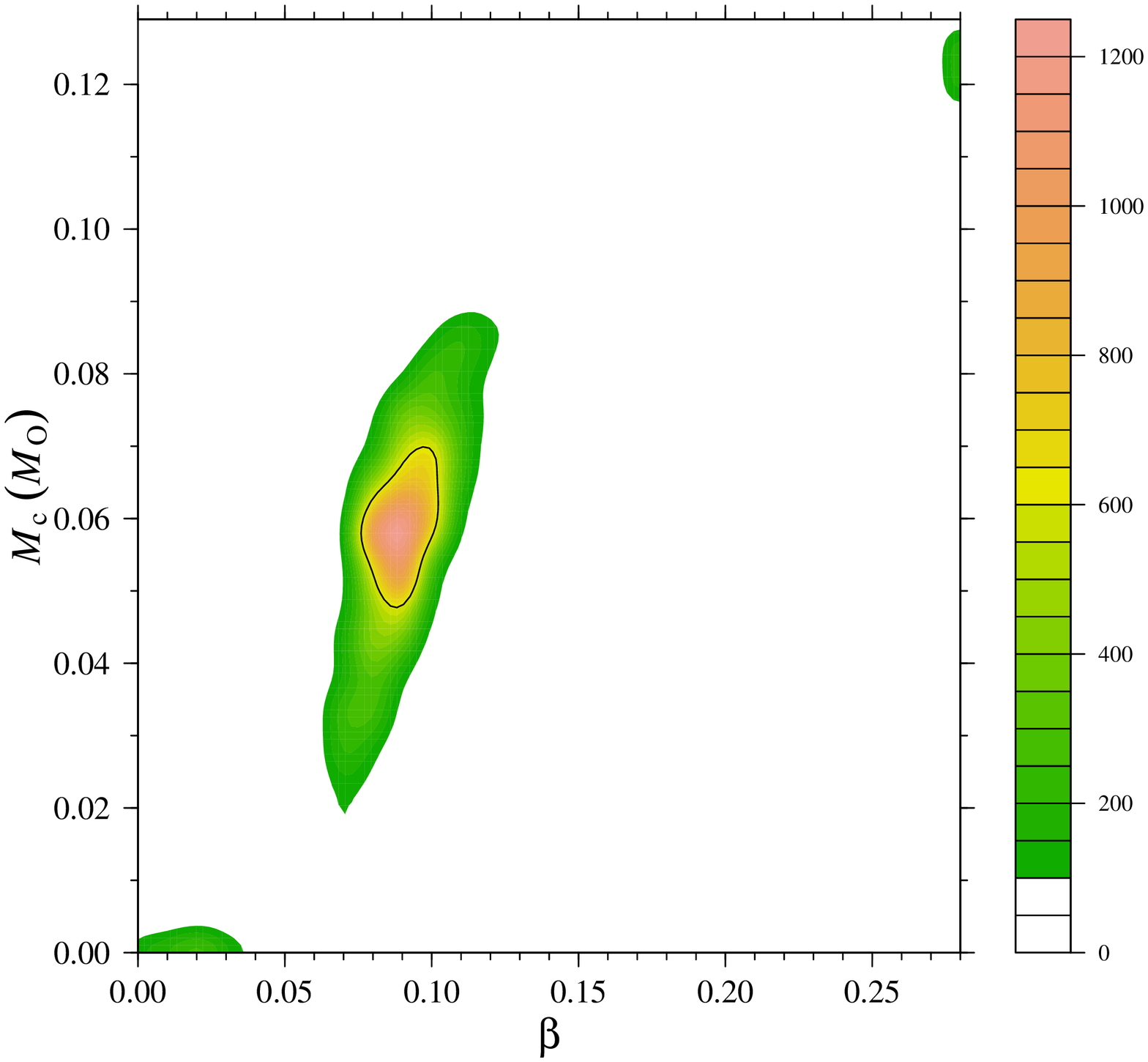}
        \includegraphics[width=8.cm,angle=-90]{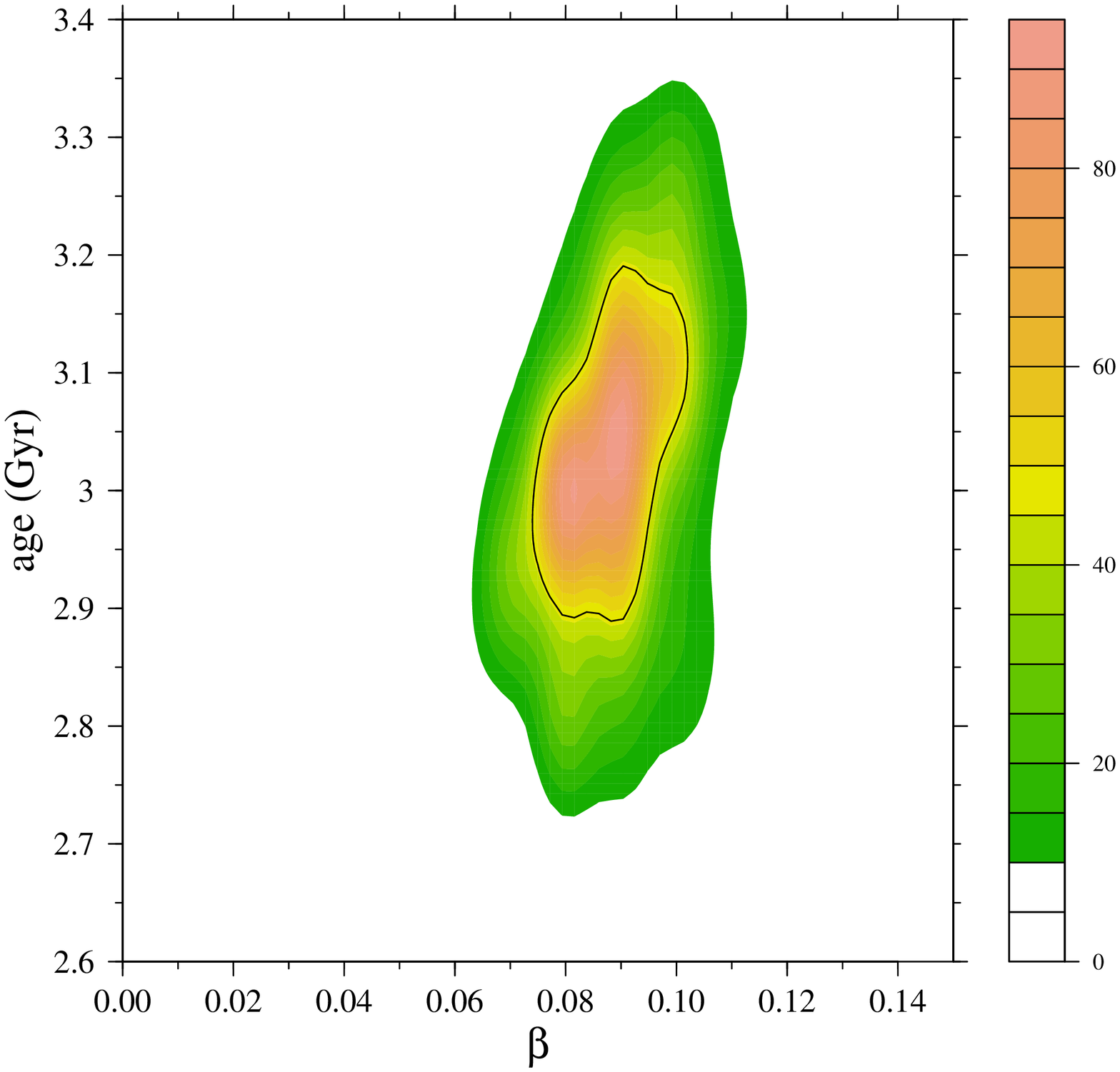}  
        \caption{Joint two-dimensional density of the probability for different couples of evolutionary parameters. {\it Left}: Joint two-dimensional density of the probability for the estimated overshooting parameter $\beta$ and the convective core mass of the binary system. The solid black line corresponds to points for which the density is half of the maximum value. {\it Right}: Same as in the left panel, but in the age versus $\beta$ plane.}
        \label{fig:ov-mcc-age}
\end{figure*}

The right panel in Fig.~\ref{fig:HR} shows the probability density for the age estimates for the whole system and for the two individual stars, which were considered independently (see Sect.~\ref{sec:method} for a detailed explanation of the differences in the fit constraints). The evolutionary stage of the two stars allows for an unambiguous fit with a single peak in the age estimation.
It is apparent that the age of the system is mainly dictated by the primary star fit; this result is in agreement with those in \citet{binary, overshooting}. The result is clearly due to the fact that the primary star is in a faster evolutionary phase than the secondary, thus allowing a better constraint on its age than the latter.

\begin{table*}
        \centering
        \caption{Result of the CPD-54 810 individual stars' fitting.}
        \label{tab:fit-sing} 
        \begin{tabular}{lccc|ccc}
                \hline\hline
                & \multicolumn{3}{c|}{primary} & \multicolumn{3}{c}{secondary} \\
                &  $q_{16}$ & $q_{50}$ & $q_{84}$ &  $q_{16}$ & $q_{50}$ & $q_{84}$\\
                \hline 
                $Y$ &  0.270 & 0.281 & 0.291 &  0.270 & 0.276 & 0.282\\
                $Z$ &  0.0104 & 0.0143 & 0.0183 & 0.0103 & 0.0128 & 0.0131 \\
                $\beta$ &  0.041 & 0.062 & 0.103 & & & \\
                $M_c$ ($M_\sun$) & 0.008 & 0.019 & 0.070 & & & \\
                age (Gyr) & 2.92 & 3.08 & 3.25 & 2.71 & 2.95 & 3.23\\
                \hline
        \end{tabular}
\end{table*}

The system fit points towards a non-negligible overshooting parameter $\beta = 0.09 \pm 0.01$. However, the meaning of this parameter is relative to the actual overshooting scheme implemented in the stellar evolutionary code. Many different choices exist in the literature, from a diffusive approach to a capped estimate in the extension of an overshooting region with respect to the core extension \citep[e.g.][]{Weiss2008, MESA2013, Paxton2018, Hidalgo2018, Nguyen2022}. The different implementations are still quite arbitrary \citep[see e.g. the discussion in][]{Salaris2017}, and therefore a better alternative is to focus on the extension of the convective core mass $M_c$. From the system fit, a value of $M_c$ of about 0.06 $M_\sun$ was obtained. The left panel in Fig.~\ref{fig:ov-mcc-age} shows the bi-dimensional density of the probability in the $M_c$ versus $\beta$ plane; as expected, there is a positive correlation between the two parameters, but -- apart two negligible islands of the solution at the grid edges -- the density shows a clear single and narrow peak. The fact that the evolutionary stage of the system is unambiguously identified in the fit is also apparent in the right panel of the figure, which shows the dependence of the estimated age on the reconstructed convective core overshooting parameter. The joint density of probability is elongated in age for a nearly constant $\beta$ parameter, suggesting that the uncertainty as to the latter is not the direct cause of the uncertainty as to the final age estimate. 

A limiting factor in the system fit is the large uncertainty as to the observed metallicity [Fe/H], which allows for a wide range of solutions at different initial metallicities. Therefore, we tested the sensitivity to this parameter by repeating the fit, halving the [Fe/H] uncertainty to 0.1 dex. The fitted system age increased to 3.12 Gyr, and this is an expected result because the preferred [Fe/H] solution in Tab.~\ref{tab:fit-bin} is slightly below $1 \sigma$ from the observational constraint. Shifting to a higher metallicity causes an increase in the age. Other changes occur in the $\beta$ parameter median value, $\beta = 0.095$, and a corresponding median core mass $M_c = 0.055$ $M_\sun$. Overall, the uncertainty as to the recovered parameters do not change, and therefore a more precise metallicity does not directly translate into more precise estimates for this particular system.

\begin{figure}
        \centering
        \includegraphics[width=8.cm,angle=-90]{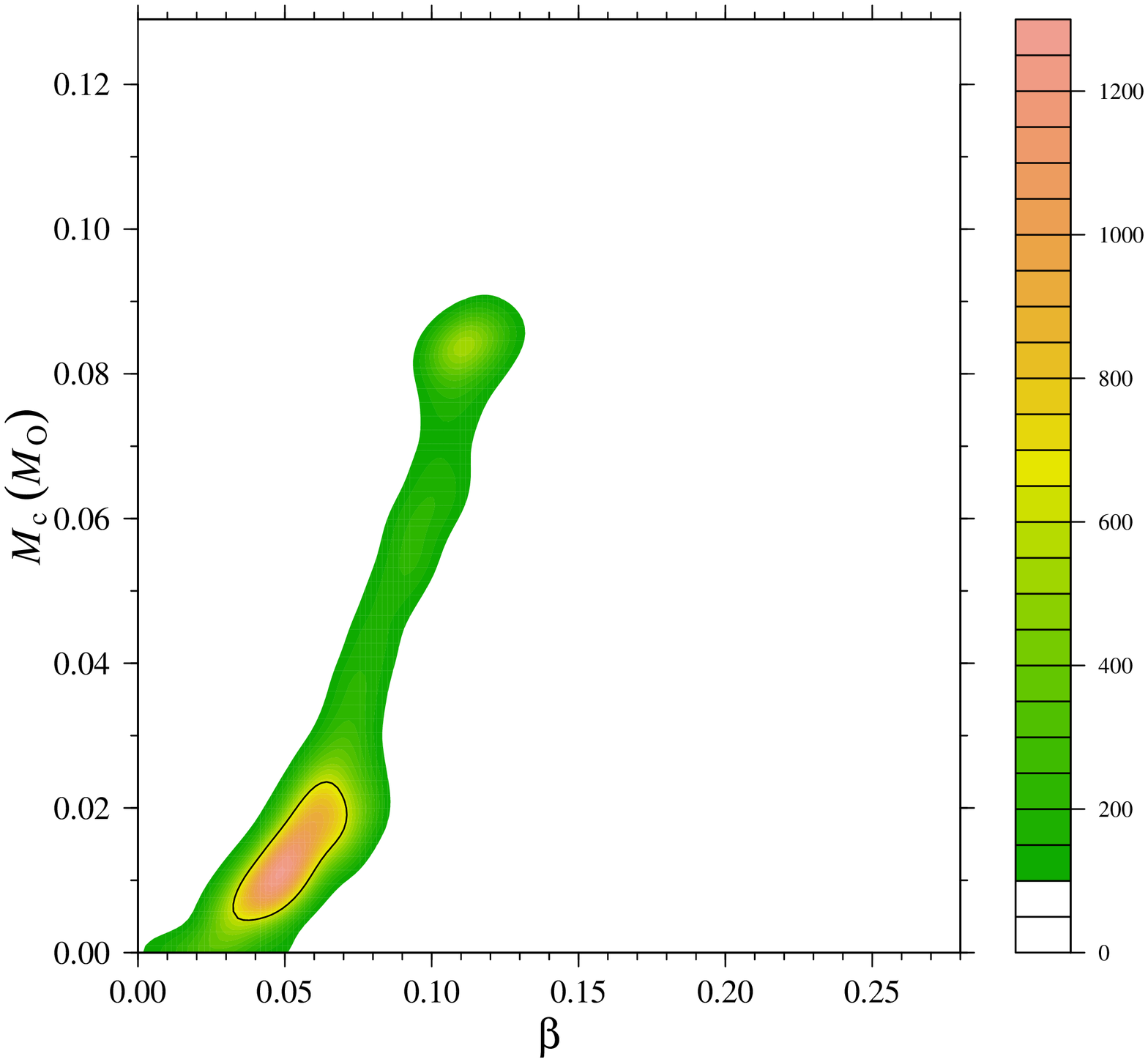}
        \caption{Same as in the left panel in Fig.~\ref{fig:ov-mcc-age}, but for the single fit of the primary star.}
        \label{fig:ov-mcc-age-prim}
\end{figure}      
      
The individual fits of the two stars show some interesting characteristics that are worth discussing. Basically the parameters inferred from them are in good agreement, and in fact the joint fit discussed above is satisfactory. However, as it is shown in Tab.~\ref{tab:fit-sing}, the initial metallicity obtained from the primary star ($Z = 0.0143$) is quite a bit higher than that from the secondary ($Z = 0.0128$), with a consequent higher estimated age for the primary star (3.08 versus 2.95 Gyr). The most interesting fact is the estimated extension of the convective core for the primary star, which is $M_c = 0.019$ $M_\sun$ and clearly lower than that obtained from the joint stars' estimate. Moreover, for the two-dimensional density of probability in the $M_c$ versus $\beta$ plane shown in Fig.~\ref{fig:ov-mcc-age-prim}, it is apparent that the median value of the estimated core is influenced by the very long tail in the distribution towards high $\beta$ and $M_c$ values, while the distribution is peaked around $M_c = 0.015$ $M_\sun$.   
The possibility to fit the primary star in rather different configurations confirms the results in \citet{binary}: the degeneracy between the initial chemical composition and the core overshooting efficiency significantly reduces the power of the analysis when the two stars are both in the MS.

\subsection{A tighter constraint in the helium-to-metal enrichment ratio}

\begin{table}
        \centering
        \caption{Result of the CPD-54 810 binary system fitting imposing the $\Delta Y/\Delta Z = 2.0$ constraint.}
        \label{tab:fit-bin-dydz2}
        \begin{tabular}{lccc}
                \hline\hline
                &  $q_{16}$ & $q_{50}$ & $q_{84}$ \\
                \hline 
                $Y$ &  0.269 & 0.272 & 0.274\\
                $Z$ &  0.0104 & 0.0116 & 0.0129\\
                $\beta$ & 0.080 & 0.090 & 0.105 \\
                $M_c$ ($M_\sun$) & 0.046 & 0.062 & 0.090 \\
                age (Gyr) &  2.94 & 3.08 & 3.25\\
                \hline
                \multicolumn{4}{c}{Fit parameters}\\
                \hline
                $T_{\rm eff,1}$ (K) &  & 6483 &  \\ 
                $T_{\rm eff,2}$ (K) &  & 6240 &  \\             
                $M_1$ ($M_{\sun}$) &  & 1.308 &   \\ 
                $M_2$ ($M_{\sun}$) &  & 1.090 &  \\ 
                $R_1$ ($R_{\sun}$) &  &  1.929 &   \\ 
                $R_2$ ($R_{\sun}$) &  & 1.181 &  \\ 
                ${\rm [Fe/H]}_1$ &  & $-0.15$ & \\
                ${\rm [Fe/H]}_2$ &  & $-0.14$ & \\      
                \hline                
                $\chi^2$ &  & 2.1 &  \\ 
                \hline
        \end{tabular}
\end{table}

The two stars in the CPD-54 810 system are too cold for a spectroscopic measurement of their helium content. Therefore, both the current and the initial helium abundance must be estimated in the fit process. This introduces a non-negligible uncertainty source, due to the concurrent interplay of metallicity, initial helium abundance, and core overshooting efficiency in setting the pace of the stellar evolution.

\begin{figure}
        \centering
        \includegraphics[width=8.cm,angle=-90]{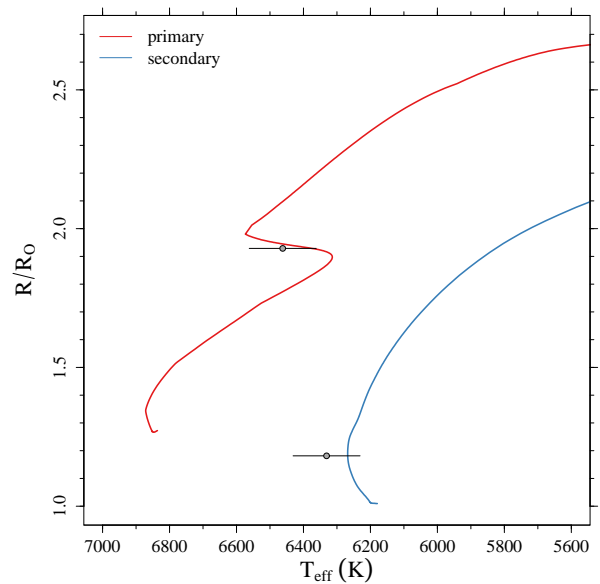} 
        \caption{Same as in the left panel of Fig.~\ref{fig:HR}, but for the results obtained adopting a fitting grid with fixed $\Delta Y/\Delta Z = 2.0$.}
        \label{fig:HR-dydz2}
\end{figure}

To fit the system, we adopted  a linear relation between initial helium abundance and initial $Z$ metallicity, a common choice among stellar modellers:  $Y = Y_p+\frac{\Delta Y}{\Delta Z} Z$. 
Adopting in the fit a grid with multiple $\Delta Y/\Delta Z$ values, the estimating pipeline proposed a value of the helium-to-metal enrichment ratio close to 2.5. This value is higher than that obtained by a recent investigation performed on the Hyades cluster \citep{Tognelli2021}. That paper significantly improved the measurement precision of this parameter, getting a value of $\Delta Y/\Delta Z = 2.03 \pm 0.15$. It is therefore interesting to  analyse the changes in the results when adopting a fitting grid with fixed $\Delta Y/\Delta Z = 2.0$.

The results, presented in Tab.~\ref{tab:fit-bin-dydz2}, show  remarkable agreement with those obtained in the previous section with a loose constraint in the helium-to-metal enrichment ratio. The pipeline is able to find a good fit also restricting the model hyperspace ($\chi^2 = 2.1$), at an age of $3.08$ Gyr (range [2.94, 3.25] Gyr), very close to the value obtained with the full grid. The same agreement holds for the convective core extension and the overshooting parameter, which are $0.062$ $M_\sun$ and 0.09, respectively. The position of the observational data with respect to the evolutionary tracks with $\Delta Y/\Delta Z = 2.0$ are shown in Fig.~\ref{fig:HR-dydz2}, where the position of best fit models are not marked for better readability but it corresponds to the points where the theoretical evolutionary tracks cross  the error bars of observational data.
   
The possibility to have a satisfactory fit in several similar configurations is partially due to the large uncertainty as to the system metallicity [Fe/H]. However, this also confirms that when both stars are in the MS, even if their masses and radii are measured with an awesome precision, they still allow for a large solution hyperspace. A system with a star in a more evolved and faster evolutionary phase would allow for a much narrower solution space, possibly highlighting some discrepancy between theoretical models and observations.  On the contrary, while the discriminating power achievable from a system with both stars in the MS is low, the estimated fundamental parameters (i.e. the age and the core overshooting parameter in the present paper) nonetheless appear to be very robust against a different assumption and constraints in the fitting process.

\subsection{Impact of the effective temperature uncertainty}\label{sec:teff}

The uncertainty as to the effective temperature reported by \citet{Miller2022} is about one half of that adopted in the previous section. 
They obtained an extremely high precision combining the measurement of the stellar radii with Gaia parallaxes, and with an accurate knowledge of
the bolometric flux \citep{Miller2020}.
As briefly mentioned in Sect.~\ref{sec:method}, our choice was mainly dictated by a cautious approach to this observational constraint. In fact, the absolute calibration of the effective temperature is still affected by large uncertainties. It is not uncommon to find different works in the literature claiming an accuracy of some tens of Kelvin; however, a  comparison of  the results by different authors on the same stars often 
show differences larger than 100 K \citep[see e.g.][]{Ramirez2005,Schmidt2016}. 

This is indeed the case for the system under investigation. The results by \citet{Ratajczak2021}, who present an analysis of this system based on a reduced data set, significantly disagree with those by \citet{Miller2022}. In fact, the former stellar effective temperatures are about 500 K cooler than the latter. Here, we are not interested in discussing the possible origins of this discrepancy, and we refer the interested reader to \citet{Miller2022}, who discuss this topic in detail. However, the large discrepancy encouraged us to adopt a cautious approach for the fit.

However, it is nonetheless interesting to analyse how the uncertainties as to the $T_{\rm eff}$ constraints propagate into the final result. Therefore, we repeated the analysis of the system (both with the full and the reduced $\Delta Y/\Delta Z = 2.0$ grids), but adopting an uncertainty of 50 K in $T_{\rm eff}$, that is, one half of that adopted in the previous fits.
This value is close to that from  \citet{Miller2022} -- that is to say 43 K -- when adding the systematic uncertainty quoted in that paper (13 K).

The results of the whole system fit with full and reduced grids are reported in Tab.~\ref{tab:fit-bin-teff}. Both fits are satisfactory with a $\chi^2 = 2.8$.
The most interesting fact is that they agree remarkably well with each other. When the algorithm is forced to refine the effective temperatures of the proposed solution at a high accuracy, the differences between the two grid solutions disappear. Moreover, the initial chemical abundances of the fits are more similar to those in Tab.~\ref{tab:fit-bin-dydz2} (for the $\Delta Y/\Delta Z = 2.0$ scenario) than to those in Tab.~\ref{tab:fit-bin}. The estimated age is 3.02 Gyr, which is the same result as that obtained with a full grid (see  Tab.~\ref{tab:fit-bin}). The extension of the convective core ($M_c \approx 0.064$ $M_\sun$) and the overshooting parameters ($\beta \approx 0.090$) agree well with the results presented above.    
In summary, an improvement in the accuracy of the effective temperatures impacts the determination of the initial chemical abundances more than the age of the system. 

\begin{table}
        \centering
        \caption{Result of the CPD-54 810 binary system fitting with a halved uncertainty as to $T_{\rm eff}$.}
        \label{tab:fit-bin-teff}
        \begin{tabular}{lccc}
                \hline\hline
                &  $q_{16}$ & $q_{50}$ & $q_{84}$ \\
                \hline 
                $Y$ &  0.269 & 0.270 & 0.272\\
                $Z$ &  0.0104 & 0.0108 & 0.0116\\
                $\beta$ & 0.090 & 0.092 & 0.100 \\
                $M_c$ ($M_\sun$) & 0.057 & 0.064 & 0.073 \\
                age (Gyr) &  2.92 & 3.02 & 3.14\\
                \hline
                \multicolumn{4}{c}{$\Delta Y/\Delta Z = 2.0$}\\
                \hline
                $Y$ &  0.269 & 0.270 & 0.272\\
                $Z$ &  0.0104 & 0.0108 & 0.0116\\
                $\beta$ & 0.090 & 0.090 & 0.100 \\
                $M_c$ ($M_\sun$) & 0.059 & 0.065 & 0.073 \\
                age (Gyr) &  2.92 & 3.01 & 3.13\\
                \hline
        \end{tabular}
\end{table}

\section{Discussion and conclusions}\label{sec:conclusions}

Profiting from the recent availability of very high accuracy observational data for the binary system CPD-54 810 \citep{Miller2022}, we attempted to fit this system to investigate the robustness of the age estimate under different assumptions. 
To do this, we used the SCEPtER pipeline \citep{scepter1, eta, binary}, a maximum likelihood procedure, on a dense grid of stellar models computed ad hoc.

Relying on the observational constraint by \citet{Miller2022}, but adopting a conservative uncertainty of 100 K in the effective temperatures, we obtained a satisfactory simultaneous fit of the system at $3.02 \pm 0.15$ Gyr. The awesome precision in the measurements of radii and masses allowed for an outstanding 5\% precision in the age estimate, which is an unusual result for MS stars. In fact the power of the investigation in MS is usually quite low \citep[see e.g.][]{binary,overshooting}.  

Taking advantage of the very precise observational data, we also tried to constrain the efficiency of the convective core overshooting in the primary star. We obtained an overshooting parameter $\beta = 0.09 \pm 0.01$, with a corresponding convective core mass $M_c = 0.059^{+0.017}_{-0.021}$ $M_\sun$.

The binary age estimate proposed in this paper is 7\% higher than the 2.83 Gyr reported by \citet{Miller2022}. This difference is at a $1 \sigma$ level or below; since \citet{Miller2022} do not report the error in the age estimate, it is impossible to precisely quantify the significance of the discrepancy.   
It is also impossible to unambiguously identify the origin of this difference because the stellar models adopted in that paper differ from those adopted here in many aspects. In particular, the treatment of the convective core overshooting is different because we did not allow it for masses below 1.1 $M_\sun$ (i.e. it does not affect the secondary star), while it was adopted even in this range for the models used in \citet{Miller2022}. Moreover, the scheme of overshooting the implementation itself is different. While we adopted a step overshooting, a diffusive approach was used in \citet{Miller2022}. While it does not directly lead to differences in the estimated age when the other model input are the same \citep[see][for a comparison of the overshooting scheme in a controlled environment]{TZFor}, it adds to other differences between the evolutionary tracks.

Another recent estimation of the system age was proposed by \citet{Ratajczak2021}, who estimated it at $3.4 - 4.0^{+0.15}_{-0.25}$
Gyr adopting PARSEC and YY models, respectively 
\citep{Yi2001, Bressan2012}. While the former estimate marginally agrees with our result at a $1 \sigma$ level, the latter is significantly higher. However, for the estimates of \citet{Ratajczak2021}, stellar effective temperatures lower by about 500 K than those adopted in our investigation were adopted. This difference forced the fit towards a higher metallicity and a lower initial helium abundance, and therefore higher ages, with respect to our models.

As opposed to previous results in the literature, we explicitly tested here the possible dependence of the estimated age on the efficiency of the convective core overshooting. Our results may help shed some light on the ongoing discussion about the dependence of the overshooting efficiency with the stellar mass \citep{Stancliffe2015, Claret2016, Constantino2018, Anders2023}. While some works suggest such a dependence exists, others do not agree with this claim. Calibrations from systems with a high precision in the radii and effective temperature measurements are of paramount importance to get further insight into this topic.

The main interest of the present work is possibly the investigation on the robustness of the age and overshooting parameter estimates. Firstly, we tested their variability when imposing a very strong prior on the helium-to-metal enrichment ratio, in practice adopting an estimation grid restricted to only stellar models computed assuming $\Delta Y/\Delta Z = 2$, which is in agreement with recent investigations \citep{Tognelli2021}. A second test was performed adopting the full grid of models, but reducing the uncertainty as to the effective temperatures to 50 K, as in \citet{Miller2022}.
The results from these different analyses were in a very good agreement with each other. The proposed age is particularly robust and only showed variations of a few tens of millions of years in the different tests performed.

While this very low variability is reassuring, it also indicates that the power of the investigation is probably low. With both stars in the MS, it was possible to find a satisfactory fit in several different configurations by only changing the initial chemical composition of the proposed solution of the system within the current uncertainty. On the other hand, this means that in a system with both stars in the MS, the estimated fundamental parameters (i.e. the age and the core overshooting parameter) appear to be very robust against a different assumption and constraints in the fitting process, contrary to systems with a star in a more evolved phase.

The robustness of the age estimate and the low uncertainty as to its value ($\approx 5\%$) should,  however, be cautiously interpreted. The results presented in this paper were obtained with a fixed grid of stellar models, all computed with identical assumptions in the input physics. A larger variability is expected when comparing results from various pipelines \citep{Reese2016, Stancliffe2016,  SilvaAguirre2017, TZFor} due to the different yet legitimate choices of stellar modellers. Ultimately, in light of the obtained results and those in the literature, a conservative yet realistic estimate of the precision achievable for a binary system age with both stars in the MS, even when their fundamental parameters are measured at very high precision, is probably about 7 - 10\%.

\begin{acknowledgements}
P.G.P.M. and S.D. acknowledge INFN (Iniziativa specifica TAsP).
\end{acknowledgements}

\bibliographystyle{aa}
\bibliography{biblio}

\end{document}